\documentstyle[12pt]{article}

\newcommand\cC{{\cal C}}
\newcommand\cH{{\cal H}}
\newcommand\cK{{\cal K}}
\newcommand\cO{{\cal O}}

\newcommand\bC{{\bf C}}
\newcommand\bH{{\bf H}}
\newcommand\bP{{\bf P}}
\newcommand\bZ{{\bf Z}}

\newtheorem{defn}{Definition}

\begin{document}

\begin{flushright}
OSU-M-95-9\\
October, 1995
\end{flushright}

\bigskip

\begin{center}
\bf \large Gromov-Witten Invariants via Algebraic Geometry
\end{center}

\bigskip

\begin{center}
Sheldon Katz \footnote{Department of Mathematics, Oklahoma State
University, Stillwater, OK 74078; katz@math.okstate.edu}
\end{center}

\bigskip

\begin{center}
\bf Abstract
\end{center}

Calculations of the number of curves on a Calabi-Yau manifold via an
instanton expansion do not always agree with what one would expect
naively.  It is explained how to account for continuous families of
instantons via deformation theory and excess intersection theory.  The
essential role played by degenerate instantons is also explained.

\section{Introduction}

In recent years, there has been much interaction between string theory
and algebraic geometry.  In particular, the Yukawa couplings for the
heterotic string compactified on a Calabi-Yau manifold $X$ can be
calculated using mirror symmetry \cite{cogp}.
The result can be interpreted as on
instanton expansion on $X$ to give the
numbers of rational curves on $X$.  Similar results hold for higher
genus \cite{bcov,bcov2}.

Certain puzzles arise from this analysis.  In examples, the effective
number of curves, or Gromov-Witten invariants, can have strange behaviors.
These numbers can be negative \cite{2param1,2param2,hkty1,hkty2,bkk},
nonintegral \cite{bcov,bcov2}, or have positive contributions
when it is geometrically clear that there are no curves of the type
naively predicted \cite{2param1,2param2,hkty1,hkty2,bkk}.

There are two techniques that are needed to resolve these puzzles.  The
first is better known.  In many situations, continuous families of instantons
can develop.  The path integral will require an integration over these
instanton moduli spaces.  This can be done in one of two ways.  The
fermion zero modes can be calculated and then integrated over as well.
This tends to be rather difficult.  The more successful approach is to
find a bundle whose Euler class calculates the virtual fundamental class
of the moduli space in question \cite{wzw}.  In practice, this may be
accomplished via deformation theory \cite{ill,laud}
and excess intersection theory \cite{ful}.
This gives rise to an obstruction analysis
and consequently an effective finite contribution of continuous families,
as illustrated for example in \cite{2param2,kontenum},
in the appendix to \cite{bcov}, or using symplectic geometry
\cite{ruan}.  The last method is based on the rigorous mathematical definition
of Gromov-Witten invariants \cite{rtop,rt}.
The second technique is prompted by the realization that certain degenerate
instantons are required in the path integral.  There are several approaches
to degenerate instantons, e.g the appendix to \cite{bcov} and
\cite{kontenum}; emphasis will be given here to the use of stable maps
as in the latter reference.

The combination of these two techniques resolves all puzzles currently posed
by calculations on the mirror theory.  There are still many calculations that
cannot presently be carried out, but to the extent that the calculations can be
performed, the results agree with those predicted by mirror symmetry.
Furthermore, it is now completely clear in which situations negative
and nonintegral contributions are possible, and these kinds of invariants
are not observed in any other cases.  Finally, in all examples for which
there are contributions to the instanton expansions yet no corresponding
curves, it is now apparent that there {\em are\/} degenerate instantons
in these cases which have a right to contribute.

In Section~\ref{count} generalities are given about counting curves and
how excess intersection theory can be used to handle continuous families
of instantons.  Several approaches to degenerate instantons are
outlined in Section~\ref{degenerate}.  Several types of examples are given
in Section~\ref{ex}: constant maps in Section~\ref{const}, families of
smooth curves in Section~\ref{smooth}, genus~0 multiple covers in
Section~\ref{multcov}, and degree~1 maps to curves of lower genus in
Section~\ref{bub}.

\section{Counting Curves}
\label{count}

\subsection{Generalities}

Consider the heterotic string theory compactified on the
Calabi-Yau target space $X$.  There are well known quantities in string
theory which are corrected by genus $g$ world sheet instantons
\cite{dsww,bcov,bcov2}, and through an instanton expansion
give methods for computing number of genus $g$ curves
on $X$ (or more precisely, the corresponding Gromov-Witten invariant).
To set up notation, let $\cK(X)\subset H^2(X,\bC)$
denote the K\"ahler cone of $X$.  Assume that $\cK(X)$ is simplicial,
generated by classes $H_1,\ldots,H_n$, where $n=\dim H^2(X)=h^{1,1}(X)$ is
the number of K\"ahler parameters.  This assumption is valid if $X$ is
an appropriate hypersurface or complete intersection in a toric variety
associated to a reflexive polyhedron \cite{batdpms,bor}.

It is desired to geometrically understand the quantity which is naively
written as
\footnote{This is not standard notation, but the notation has been chosen to
emphasize the similarities between the various genera.}
\begin{equation}
\label{Fg}
F_g=\sum_{C_g\subset X} q_1^{C_g\cdot H_1}\cdots q_n^{C_g\cdot H_n}
\end{equation}
where the sum is over genus $g$ (possibly degenerate) holomorphic instantons.

Here $\omega=B+iJ$ is the complexified K\"ahler class ($B$ field and metric),
and in terms of the chosen basis for $\cK(X)$, put
\begin{equation}
\label{qs}
\omega=\sum_{i=1}^n t_i H_i\\
q_i=e^{2\pi i t_i}
\end{equation}
The formula~(\ref{Fg}) for $F_g$ must be understood to contain terms
associated to {\em degenerate
instantons\/}, including what is often called the classical term (which
results from integration over the space of constant maps from the world
sheet).

Typically, $F_g$ is calculated via mirror symmetry, and then geometric
information can be inferred from the result inductively using the quantities
$F_{g'}$ for $g'<g$ and an understanding of certain degenerate instantons.

The quantity $F_0$ is just the prepotential.  The quantity $F_1$ is introduced
in \cite{bcov}, while $F_g$ for $g\ge 2$ is introduced in \cite{bcov2}.

Put $\delta_i = q_i \frac{\partial}{\partial q_i}$.  Then the third derivative
$\delta_i \delta_j \delta_k F_0 $ is a Yukawa coupling, which is calculated
to be\footnote{This is still a naive formula; the role of
degenerate instantons will be
clarified later.}
\begin{equation}
\label{g0}
\sum_{C\subset X}
(C\cdot H_i)(C\cdot H_J)(C\cdot H_k)
q_1^{C\cdot H_1}\cdots q_n^{C\cdot H_n}
\end{equation}
where the sum is over genus~0 instantons.

For ease of notation, if $[C]=\gamma\in H^2(X,\bZ)$, put
\begin{equation}
\label{multi}
q^\gamma=\prod_{i=1}^n q_i^{C\cdot H_i}
\end{equation}
Further, put $d_i=C\cdot H_i$ whenever the meaning of $C$ is clear from
context.
Let $n_\gamma$ be the number of genus~0 curves $C$ with $[C]=\gamma$.
The expected dimension of the space of genus~0 curves (or any genus for that
matter) is 0, so it makes sense to seek the number of such curves, counted
appropriately.
Then $\delta_i \delta_j \delta_k F_0 $ can be rewritten as
\begin{equation}
\sum_\gamma n_\gamma d_id_jd_kq^\gamma+\ldots =
\sum_{f: P^1\to X, f(p_\alpha)\in H_\alpha} q^\gamma + \ldots,
  \label{yukawa}
\end{equation}
where the $p_\alpha\in\bP^1$ are marked points and the dots denote
terms arising from degenerate instantons.
It is thus seen that $n_\gamma$ is (essentially) a Gromov-Witten invariant.
\footnote{There are several different proposed definitions of the Gromov-Witten
invariant; it is expected that they all agree.}

Similarly, the quantity $\delta_i F_1$ allows for enumeration of genus~1
curves, and is seen to be
\begin{equation}
\sum_\gamma e_\gamma d_iq^\gamma+\ldots =
\sum_{f:E\to X, f(p_i)\in H_i} q^\gamma +\ldots.
  \label{g1}
\end{equation}
where $E$ is any elliptic curve and $e_\gamma$ is the number of genus~1
instantons.  This will again be corrected for degenerate instantons later.

The mathematical reason why it is necessary to consider 3~point functions
and 1~point functions in (\ref{yukawa}) and (\ref{g1}) above is that
the space of automorphisms of a genus $g$ curve has respective dimensions~3
and~1 for $g=0$ and 1, so rigidification is necessary to get finite numbers
of curves.

\label{general}

\subsection{Excess Intersection}
It is not possible in general to get finite instanton moduli spaces, even
after rigidification as above in theories with generic parameter values.

To see what to do, first consider a special case.  Suppose that the
instanton moduli space is infinite yet
smooth of finite dimension for a special parameter
value.  Suppose that the moduli space becomes finite for general parameter
values.

Deformation theory \cite{ill,laud} gives rise to an obstruction bundle on the
instanton moduli space.
As the special parameter value gets smoothly
perturbed to a general
parameter value, a section of the obstruction bundle results, and
the limits of the finite moduli spaces are recognizable
as the zero locus of this section of the obstruction bundle.
The number of zeros is just the degree of the top Chern class (or Euler
class) of the obstruction bundle.

Notice that the form of the result is independent of the assumption that
the moduli space becomes finite for general parameter
values.  So it is natural to
make the hypothesis that the effective finite contribution
of our moduli space is just the degree of the top Chern class of the
obstruction bundle.

In string theory, the calculation is performed by an integration over the
fermion zero modes; it is being asserted that
the Chern class calculation achieves the
same result.

This assertion can be justified even in situations where the moduli
space is infinite for generic parameter values.  The idea is to interpret
our calculation in a more general context, so that what would be called
general parameter values in the original context becomes special in the
more general context.  For example, the calculation of the Yukawa couplings
can be understood as an A-model calculation.  The A-model makes sense for
almost complex manifolds \cite{tsm}.   The moduli
space of rigidified pseudoholomorphic curves on a generic almost complex
manifold is finite, and the assertion is reduced to the case already
checked. An example is given in \cite{2param2}.

There is a more general method which will merely be mentioned here.
The above procedure is a special case of {\em excess intersection
theory\/} \cite{ful}.  In excess intersection theory, there are not
only bundles but a certain subcone of these bundles, the {\em normal
cone\/}.  In fact, a calculation in \cite{wzw}
is recognizable as excess intersection theory.  This led the author in
\cite{Dyrkolbotn}
to define an analogous {\em virtual normal cone\/} in the special
case that the reduced moduli space is smooth.  The resulting effective
finite contribution was called the {\em equivalence\/} following the
corresponding terminology in excess intersection theory \cite{ful}.
Other authors
refer to the same object as the {\em virtual fundamental class\/}.
J.~Li and G.~Tian \cite{litian}
have recently found a good definition of the virtual normal
cone in the general case.  Work is in progress to compare this definition
with the existing definition \cite{rtop,rt} using symplectic geometry.

\section{Degenerate Instantons}
\label{degenerate}

Degenerate instantons first appeared in \cite{aspmor}
in the course of compactifying
the instanton moduli space of multiple cover maps to a rational curve.  On
the other hand, the
fact that the instantons were degenerate played no role in the calculation.

It became clear from \cite{bcov} that degenerate
instantons play a fundamental role; the path integral can reduce
to a space on which all instantons are degenerate.  This occurs for degenerate
maps of degree~1 from an elliptic curve to a rational curve.

One approach is to identify instantons $f:C\to X$ with their graphs
$\Gamma\subset C\times X$, and then consider degenerate configurations
of $\Gamma$ which do not necessarily arise from maps $f$.  This is the
approach taken in \cite{aspmor} and the appendix to \cite{bcov}.  This space of
degenerate instantons is a subset of the {\em Hilbert scheme\/} \cite{groth}
of $C\times X$, (or a relative Hilbert scheme, if $C$ can vary).

Another approach, and the one that will be expanded on in the sequel, is to
consider stable maps as the degenerate instantons.

One should ask whether these methods of compactification are compatible.
{}From the point of view of the path integral, it is not a complete surprise
that different methods of compactifying the instanton moduli space give
the same answers.  We can view the definition of the path integral as
an integral over the interior of the moduli space of all maps.
If part of the path integral is concentrated near the
boundary of the moduli space of all maps, then the method of compactification
should not affect the result of the calculation, just the way that the
calculation proceeds.

On the other hand, from the viewpoint of deformation theory,
a very precise method of calculation has been proposed
which ignores the interior of the
moduli space, so there is no apparent reason why the methods are equivalent.
Yet there are no discrepancies evident to date between different methods.

Kontsevich introduced the notion of a stable map as follows \cite{kontenum}.

\begin{defn}
A stable map $f:C\to X$ consists of the following data.
\begin{itemize}
\item A curve $C=C_1\cup \ldots\cup C_n$ whose only singularities
are ordinary double points.
\item Marked points $p_1,\ldots,p_r\in C$ in the smooth locus of $C$.
\item A holomorphic map $f:C\to X$ such that there are no continuous
automorphisms of $f$ which are the identity on $p_1\ldots,p_r$.
\end{itemize}
\end{defn}
The last condition means that smooth components $C_i$ of genus 0
(resp.~1) have at least 3 (resp.~1)special (i.e.\ nodal or marked)
points.

For counting curves, it is necessary to take
\begin{equation}
  r=\left\{ \begin{tabular}{cl}
            0 & $g\ge 2$\\
            1 & $g=1$\\
            3 & $g=0$
  \end{tabular} \right.
  \label{r}
\end{equation}
This choice has been explained in the last paragraph of
Section~\ref{general}.

One early impetus to the study of degenerate instantons arose in conversations
between the author and D.~Morrison in~1991.  It was realized that
during a change of complex structure of the quintic threefold $\bP^4[5]$,
a conic can approach a line pair.  Suppose that the line pair occurs for
$t=0$, while smooth conics occur for $t\neq 0$.
For $t\neq 0$ we have maps
from $\bP^1$ with $\gamma\cdot H=2$ ($H$=hyperplane class), but no such
maps for $t=0$.  Several strategies for including degenerate instantons
were proposed to explain this discontinuity.

Stable maps provide the most straightforward resolution of this problem.
More generally, suppose that a smooth rational curve $D$ degenerates
to a union $D_1\cup D_2$ of two smooth rational curves, meeting
transversally at a point.

Now count rigidified stable maps whose image is $D$, and compare to those
whose image is $D_1\cup D_2$.  The contribution of $D$ to a Yukawa coupling
$\langle H_1H_2H_3 \rangle$ is
\begin{equation}
(D\cdot H_1)(D\cdot H_2)(D\cdot H_3)
  \label{smoothcont}
\end{equation}

For $D_1\cup D_2$ there are several types of contributions.  The source
$C$ of the stable map must be a union $C_1\cup C_2$ of
two $\bP^1$s.  The three marked
points can be distributed arbitrarily between the curves $C_i$.  Then
each of the curves $C_i$ can map to either $D_1$ or $D_2$.
Including all possibilities, the
calculation becomes
\begin{eqnarray}
\prod_i(D_1\cdot H_i)+\prod_i(D_2\cdot H_i) + \nonumber\\
\sum_{\{a,b,c\}=\{1,2,3\}}\left[(D_1\cdot H_a)(D_1\cdot H_b)(D_2\cdot H_c)+
(D_1\cdot H_a)(D_2\cdot H_b)(D_2\cdot H_c)\right]
  \label{reduced}
\end{eqnarray}
Using the relation
\begin{equation}
D\cdot H_i=(D_1+D_2)\cdot H_i,
  \label{degree}
\end{equation}
it is clear that the results of (\ref{smoothcont}) and (\ref{reduced})
agree.

\section{Examples}
\label{ex}
In this section, examples will be given for the
use of stable maps as they arise in the instanton expansion for $F_g$.

\subsection{Constant Maps}
\label{const}
The first application is to the classical term, which corresponds to
constant maps.

\subsubsection{Genus~0}
In this case, 3~marked points are needed.  For a constant map with 3~marked
points to be stable, the source curve $C$ must be irreducible, hence just
$\bP^1$.  More generally, a constant map is stable if and only if the
source of the map is a stable curve.
The three marked points may be assumed to be
$0,1,\infty\in \bP^1$.  The only other data needed to describe the map
is the image point $p\in X$.  Thus the moduli space is isomorphic to $X$.
The dimension is~3, so the obstruction bundle is trivial.  The point
condition given by $H_i$ translates into the cohomology class of $H_i$ on
$X$.  This gives \cite{strwit}
\begin{equation}
\delta_i\delta_j\delta_kF_0=H_iH_jH_k+\ldots
  \label{constmap0}
\end{equation}

\subsubsection{Genus~1}

A single marked point $p\in C$ is needed in this case.  As noted above,
the source of the map is a stable curve.
Pointed stable curves $p\in C$ are parameterized by $\overline{M_{1,1}}$.
Including the target point $p\in X$ gives the moduli space
$\overline{M_{1,1}}\times X$.  This space has dimension~4, hence there is a
rank~3 obstruction bundle whose Chern class must be calculated.
For fixed $C$,
the obstruction space is given by the second hypercohomology group $\bH^2$
of the
complex\footnote{This calculation is only valid when $C$ is irreducible, but
it can be shown that the result is the same in any case.}
\begin{equation}
T_C(-p)\to f^*T_X
  \label{complex}
\end{equation}
where $T_C(-p)$ denotes the sheaf of vector fields on $C$ which vanish
at $p$ \cite{kontenum}.  The resulting cohomology sequence shows that the
obstruction space is just $H^1(f^*T_X)$.  Since $f^*(T_X)$ is a trivial
bundle, this gives $\bH^2\simeq H^1(\cO_X^3)$.
This globalizes to the rank~3 bundle
\begin{equation}
\cH^*\otimes T_X,
  \label{obundle}
\end{equation}
a bundle on $\overline{M_{1,1}}\times X$.  Here $\cH$ is the Hodge bundle of
holomorphic 1~forms on $C$.  The Hodge bundle arises because of the isomorphism
$H^1(\cO_C)\simeq (H^{1,0}(C))^*$.

Inserting the condition $p\in H_i$ gives
\begin{equation}
\begin{tabular}{cc}
$\int_{\overline{M_{1,1}}\times X} c_3(\cH^*\otimes T_X)\cdot H_i$ & $=$\\
$\left[\int_X c_2(T_X) \cdot H_i \right] \int_{\overline{M_{1,1}}} c_1 (\cH^*)$
      & $=$ \\
$-\frac1{12}\int_X c_2(T_X)\cdot H_i$ &
  \label{constantmap1}
\end{tabular}
\end{equation}
the constant term of $\delta_i F_1$ \cite{bcov}

\subsubsection{Genus $g\ge 2$}

In this case, there are no marked points, and the
parameter space is $\overline{M_g}\times X$ which has dimension~$3g$.

If $f:C\to p\in X$ is a constant map, then the obstruction space is
again seen to be $\cH^*\otimes T_X$, a rank~$3g$ bundle
on $\overline{M_g}\times X$.
So the $\gamma=0$ contribution to $F_g$ is
\begin{equation}
\int_{\overline{M_g}\times X} c_{3g}(\cH^*\otimes T_X) =
(-1)^g\frac12 \chi(X)\int_{\overline{M_g}} c_{g-1}(\cH)^3
  \label{constmaps}
\end{equation}
agreeing with \cite{bcov2} up to sign.

\subsection{Families of Smooth Curves}
\label{smooth}

Suppose that $X$ contains is a family of smooth curves with smooth
parameter space $B$.  Denoting the total space of the family by $\cC$,
there results a diagram
\begin{equation}
  \begin{tabular}{ccc}
$\cC$ &$ \subset$ &$ B\times X$\\
 $ \pi\downarrow$  &  & $\downarrow$\\
$B$    & $=$         &$ B$
  \end{tabular}
  \label{family}
\end{equation}
The fibers $\pi^{-1}(b)=C_b\subset \{b\} \times X$ are identified with curves
$C_b\subset X$.
Since the curves are smooth, the point conditions may be inserted at the
outset of the analysis for genus $g<2$.  This reduces the deformation problem
from the study of maps to the study of embedded curves.
Let $N_b$ denote the normal bundle of $C_b\subset X$.  Then
\begin{equation}
  \begin{tabular}{ccl}
$H^0(N)$ & $=$ & $T_bB$ \\
$H^1(N)$ & $=$ & obstruction space
  \end{tabular}
  \label{Ti}
\end{equation}
Note that $H^1(N)$ is dual to $H^0(N)$ via the pairing
\begin{equation}
H^1(N)\otimes H^0(N)\to H^1(\wedge^2N)\simeq H^1(T_C^*)=\bC
  \label{pairing}
\end{equation}
so the obstruction bundle is globally $T_B^*$.

It follows that the contribution to the Gromov-Witten invariant is
\begin{equation}
c_{\dim(B)}(T_B^*)=(-1)^{\dim(B)}\chi(B)
  \label{equiv}
\end{equation}

\noindent
{\em Examples\/}.

\noindent
1. For a ruled surface $E\subset X$:
\begin{equation}
  \begin{tabular}{ccc}
$E$ & $\subset$ & $X$\\
$\pi\downarrow$ & & \\
$B$ & &
  \end{tabular}
  \label{ruled}
\end{equation}
we get the contribution $2g-2$, where $g$ is the genus of the parameter
curve $B$.

Note that if $g(B)\ge 1$, the complex structure on $X$ is not generic:
deforming the complex structure of $X$ destroys $E$, replacing
it with $2g-2$ isolated rational curves \cite{2param1}.

\noindent
2. If $\bP^2\subset X$, then lines in $\bP^2$ are parameterized by a
$\bP^2$, hence the contribution to the Gromov-Witten invariant is 3.
Plane conics are parameterized by $\bP^5$, hence the contribution to
the Gromov-Witten invariant is $-6$.  In fact, a deformation of
almost complex structure reveals a finite number of pseudoholomorphic
oriented curves, with signed number $-6$ \cite{ruan,2param2}.

Without exception, these check out against calculations performed using
mirror symmetry \cite{2param1,2param2,hkty1,hkty2,bkk}.

\subsection{Genus~0 Multiple Covers}
\label{multcov}

Degree $k$ multiple covers of a smooth curve $C\subset X$ with
$[C]=\gamma$ contribute
$d_id_jd_kq^\gamma$ to $\delta_i\delta_j\delta_k F_0$
\cite{cogp,aspmor}.  As an alternative approach, the space of instantons can be
compactified in two ways: using the normal bundle of graphs of maps or stable
maps.  Using the normal bundle of maps gives rise to the same moduli
space as in \cite{aspmor} but a different obstruction bundle.  The
result is that all answers agree, giving for
$\delta_i\delta_j\delta_k F_0$
\begin{equation}
H_iH_jH_k+\sum_\gamma\frac{d_id_jd_kn_\gamma q^\gamma}{1-q^\gamma}
  \label{multcover}
\end{equation}

The use of stable maps explains a phenomenon which was not previously
understood.  In all cases where there is a map $\bP^1\to X$ which is not
an embedding, there are contributions to $n_{k\gamma}$ which arise
in addition to the usual contributions for multiple covers.  Up until now,
this was only realized by the calculation of Yukawa couplings via mirror
symmetry.

\noindent
{\em Example\/}: Consider a nodal rational curve $D$.  Such a curve is
the image of an immersion $g:\bP^1\to X$ which maps two points of
$\bP^1$ to the node $p$ and is otherwise an embedding.  The curve $D$ has
two branches near $p$.
The multiple covers discussed above arise in this situation by
composing $g$ with a multiple
cover $\bP^1\to\bP^1$.

In addition, other $k$-fold covers of $D$ arise as follows.  Let
$C=C_1\cup\ldots\cup C_k$ be a union of $k$ copies of $\bP^1$ meeting
transversely, with $C_i\cap C_{i+1}=p_i$ for $i=1.\ldots,k-1$.
Form a map $f:C\to D$ by setting the restriction of $f$ to each $C_i$ to
be the immersion $g$ above.  In addition, it is
required that  $f(p_i)=p$ for all $i$, and then
further required that for each $i$, the branches of $D$ to which complex
disks around $p_i$ in $C_i$ and $C_{i+1}$ are mapped are not the same
branch.
This phenomenon occurs in many examples, e.g.\ \cite{2param2,hkty1,hkty2,bkk}.

\noindent
{\em Example\/}: Union of a line $L$ and a conic $D$ inside a
$\bP^2$ contained in
$X$.  Here there are two intersection points $q_1,q_2$
with two branches at each point.
For simplicity, rather than talking about the homology class $\gamma$,
the planar degree will be described instead.

The construction of multiple covers is similar to that above.  Multiple
cover maps do not come in arbitrary degree; the degree can only be added
to in multiples of~3.  For example, to
get a degree~4 degenerate instanton from a degree~1 instanton:
start with an instanton $f:C_1\to L$.  Put $C=C_1\cup C_2\cup C_3$ where
the $C_i$ are rational curves and $C_i\cap C_{i+1}=p_i$ as before.
Extend $f$ to $\tilde{f}:C\to L\cup D$ by reparameterizing $f$ as necessary
so that $f(p_1)=q_1$ (or $q_2$, which is similar), then define $\tilde{f}$
on $C_2$ so that $\tilde{f}$ is an isomorphism of $C_2$ onto $D$ with
$\tilde{f}(p_2)=q_2$, and finally defining $\tilde{f}$ to be an isomorphism
of $C_3$ onto $L$.

\subsection{Degree~1 maps to curves of lower genus}
\label{bub}
This is the ``bubbling phenomenon'' of \cite{saksuhl,gromov,bcov,bcov2}.
Consider the contribution
$e_{g,g'}$ of genus $g'$ curves $C$ to $F_g$ with $g'<g$.  For ease of
notation, put $g''=g-g'$.  These degenerate instantons
arise from stable
maps $f:C_1\cup C_2\to C$ with $C_1,\ C_2$ of respective genera
$g'',\ g'$ such that $f(C_1)$ is a point $p\in C$ and $f$ is an isomorphism
from $C_2$ to $C$.
Restrict attention to the case $g\ge 2$ to avoid having
to say anything about the point conditions
(these do not create any
new problems).
The moduli space of such stable maps is
\begin{equation}
C\times\overline{M_{g'',1}},\quad\dim=3g''-1
  \label{bubble}
\end{equation}
It is very desirable to work out the obstruction theory
for this deformation problem.  This has not yet been done.

The contribution $e_{1,0}=1/6$ has been worked out mathematically in the
appendix to \cite{bcov} using degenerate graphs as degenerate instantons; there
the obstruction analysis is simpler than in the problem just outlined.
In this case as well as the case of genus~0 multiple covers considered
in Section~\ref{multcover} the use of stable maps seems to complicate
the analysis.

In \cite{bcov2}, the values $e_{2,0}=1/240$ and $e_{2,1}=0$ where calculated
by the method of undetermined coefficients in a calculation of $F_2$ for
the quintic threefold with the aid of mirror symmetry.

There is a bit of speculation worth making here.  Preliminary analysis
makes it conceivable
that the obstruction problem on $C\times\overline{M_{g'',1}}$ factors
naturally into a complicated term on $\overline{M_{g'',1}}$ and the tangent
bundle of $C$.  This would imply that for $g-g'\ge 2$
\begin{equation}
e_{g,g'}=(1-g')e_{g-g',0}
  \label{reduction}
\end{equation}
Note that this is consistent with $e_{2,1}=0$, and would imply that
$e_{g,1}=0$ for arbitrary $g$ (this last equality is also consistent with the
reasoning in \cite{bcov2}).  The first non-trivial check of this speculation
would be to see if $e_{4,2}=-1/240$.

\section{Conclusions}
It appears that any natural method in algebraic geometry for compactifying
instanton moduli spaces and  performing the path integral over fermion zero
modes gives answers which agree with calculations done by mirror symmetry.
Deformation theory provides a natural framework for the analysis.
The incorporation of degenerate instantons appears essential in any formulation
of the theory.

\section{Acknowledgements}  The author wishes to thank C.~Faber, D.R.~Morrison,
and C.~Vafa for helpful conversations at various points during which some
of these ideas were being formulated.
This research was supported in part by NSF grant DMS-9311386.

\end{document}